\documentclass[preprint]{imsart}
\usepackage{amsthm,amsmath,amssymb,amsfonts}

\usepackage{array}
\usepackage{graphicx}

\usepackage{pgfplots}
\usepackage{color}
\usepackage[english,ruled,vlined]{algorithm2e}

\usepackage[T1]{fontenc}

\usepackage{textcomp}

\newcommand{\argmax}{\textup{argmax}}

\newtheorem{theorem}{Theorem}
\newtheorem{definition}{Definition}

\RequirePackage[round]{natbib}
\pgfmathdeclarefunction{gauss}{2}{%
	\pgfmathparse{1/(#2*sqrt(2*pi))*exp(-((x-#1)^2)/(2*#2^2))}%
}

\lastpage{0}

\startlocaldefs
\endlocaldefs

\begin{document}

\begin{frontmatter}

\title{\textsc{A Parsimonious Tour \\ of Bayesian Model Uncertainty}}
\runtitle{\textsc{Bayesian Model Uncertainty}}

 \author{\fnms{\textsc{Pierre-Alexandre}} \snm{\textsc{Mattei}}\ead[label=e1]{pierre-alexandre.mattei@inria.fr}}
 \address{Universit\'e C\^ote d'Azur\\ Inria, Maasai project-team \\ Laboratoire J.A. Dieudonn\'e, UMR CNRS 7351 \\ \printead{e1}}


\runauthor{\textsc{Mattei}}

\begin{abstract}
	
	Modern statistical software and machine learning libraries are enabling semi-automated statistical inference. Within this context, it appears easier and easier to try and fit many models to the data at hand, thereby reversing the Fisherian way of conducting science by collecting data \emph{after} the scientific hypothesis (and hence the model) has been determined. The renewed goal of the statistician becomes to help the practitioner choose within such large and heterogeneous families of models, a task known as \emph{model selection}. The Bayesian paradigm offers a systematized way of assessing this problem. This approach, launched by Harold Jeffreys in his 1935 book \emph{Theory of Probability}, has witnessed a remarkable evolution in the last decades, that has brought about several new theoretical and methodological advances. Some of these recent developments are the focus of this survey, which tries to present a unifying perspective on work carried out by different communities. In particular, we focus on non-asymptotic out-of-sample performance of Bayesian model selection and averaging techniques, and draw connections with penalized maximum likelihood. We also  describe recent extensions to wider classes of probabilistic frameworks including high-dimensional, unidentifiable, or likelihood-free models.
\end{abstract}



\end{frontmatter}

\tableofcontents

\section{Introduction: collecting data, fitting many models}

Today, the conventional statistical process embodied by Fisher's \citeyearpar{fisher1938} famous exhortation  
\begin{quote} 
	\emph{To consult the statistician after an experiment is finished is often merely to ask him to conduct a post mortem examination. He can perhaps say what the experiment died of.}
\end{quote}
has been somewhat reversed. Indeed, as illustrated e.g. by \citet[Chapter 1]{giraud2014introduction} or by \citet{cox2018big}, modern scientific research often involves the simultaneous measurement of a large number of (potentially irrelevant) variables before statistical practice is actually set in motion. Rather than falsifying or corroborating predetermined hypotheses, researchers mine these high-dimensional data using a vast toolbox of statistical models. This new scientific practice---caricatured by the motto \emph{collect data first, ask questions later}---was notably powered by the recent rise of automatic statistical software, illustrated for example by the growing popularity of Stan \citep{carpenter2016}, PyMC \citep{salvatier2016}, or JASP \citep{love2019jasp},.

In this new context, it appears of paramount importance to be able to compare and assess the relevance and the performance of these many models, and to identify the best ones. \emph{Bayesian model uncertainty} provides a systematized way of answering some of these questions. This approach, whose history is briefly summarized in the next section, has witnessed a remarkable evolution in the last decades, that has brought about several new theoretical and methodological advances. The foundations of Bayesian model uncertainty, as well as some of these recent developments are the subject of this review paper. In particular we focus on new perspectives on the out-of-sample properties of Bayesian model selection and averaging. We also review computational advances that enable practitioners to assess model uncertainty in three challenging families of modern statistical models: singular, high-dimensional, and likelihood-free models.

\section{A brief history of Bayesian model uncertainty}

Bayesian model uncertainty is essentially founded on the idea of spreading prior beliefs between competing models, implying that the marginal distribution of the data follows a mixture of all model-specific marginals. This paradigm was initially developed by Sir Harold Jeffreys and Dorothy Wrinch in a series of papers \citep{wrinch1919,wrinch1921,wrinch1923}, culminating with Jeffreys's book \emph{Theory of Probability} \citeyearpar{jeffreys1939}. For a recent perspective on the place of Bayesian model uncertainty in \emph{Theory of Probability}, see \cite{robert2009}. It is worth mentioning that Jeffreys considered it an essential piece of his scientific work. Indeed, in a 1983 interview with Dennis Lindley quoted by \citet{etz2017}, Jeffreys stated that he thought that his most important contribution to probability and statistics was ``the idea of a significance test (...) putting half the probability into a constant being $0$, and distributing the other half over a range of possible values''.

Independently, similar ideas were developed by J. B. S. \cite{haldane1932}, as recently exhibited by \citet{etz2017}, and also by Alan Turing, who designed related tests to decrypt Enigma codes during World War II, as testified by Turing's main statistical collaborator I. J. \cite{good1979}.

This scientific paradigm gained considerable popularity in the beginning of the 1990s, in particular with David MacKay's \citeyearpar{mackay1991} thesis which had a significant impact on the then-burgeoning machine learning community, and with the review paper of Robert Kass and Adrian Raftery \citeyearpar{kass1995}, which quickly disseminated Jeffreys's ideas to the whole scientific community.

\section{The foundations of Bayesian model uncertainty}

In this section, we present the Bayesian framework of model uncertainty, essentially founded by Jeffreys in his book \emph{Theory of Probability} \citeyearpar{jeffreys1939}. We start with some data $\mathcal{D}$ living in a probability space and with a family $\mathcal{M}_1,...,\mathcal{M}_{d_{\textup{max}}}$ of candidate statistical models. Unless specified otherwise, these models correspond to parametric families (indexed by $\Theta_1,...,\Theta_{d_{\textup{max}}}$) of probability measures over the data space, which are absolutely continuous with respect to a reference measure (usually the Lebesgue or the counting measure).

\subsection{Handling model uncertainty with Bayes's theorem}

The Bayesian framework may be summarized in a single sentence: \emph{model unknown quantities as random variables in order to assess their uncertain nature}. Under model uncertainty, there are two different kinds of unknowns: models and their parameters. We assume therefore that priors $p(\mathcal{M}_d)$ and $p(\theta | \mathcal{M}_d)$ over these unknowns are specified. As in \cite{draper1995}, we may summarize this framework by contemplating what we will call the \emph{expanded model}

\begin{equation}
\mathcal{M}_{\textup{Exp}}:\left\{
\begin{array}{ll}
\mathcal{M}_d \sim p(\cdot) \\
\theta \sim p(\cdot | \mathcal{M}_d) \\
\mathcal{D} \sim p(\cdot|\theta, \mathcal{M}_d).
\end{array}
\right.
\end{equation}
A way of interpreting this three-stage hierarchical model is to see the global prior distribution (over both models and parameters) as a way of sampling distributions $p(\cdot|\theta, \mathcal{M}_d)$ over the data space. In this very general framework, it is actually not necessary to assume that the model family is finite, or even countable. \cite{draper1995} advocates for example the use of a continuous model family to gain flexibility, and shows several applications (see also \citealp[Chapter 7]{gelman2014}). Note that the resulting marginal distribution of the data is a mixture model (an infinite one in the case of an infinite model family).

Now that we have specified the probabilistic architecture, model uncertainty will be tackled automatically by the Bayesian machinery.
Indeed, from Bayes's theorem, we obtain posterior probabilities of models as, for all $d \in \{1,...,d_{\textup{max}}\}$,
$$p(\mathcal{M}_d|\mathcal{D}) \propto p(\mathcal{D}|\mathcal{M}_d) p(\mathcal{M}_d)  ,$$
where 
\begin{equation}
p(\mathcal{D}|\mathcal{M}_d) = \int_{\Theta_d} p(\mathcal{D}|\theta,\mathcal{M}_d)p(\theta|\mathcal{M}_d)d\theta
\end{equation}
is the \emph{marginal likelihood} of model $\mathcal{M}_d$---also called \emph{evidence} (see e.g. \citealp{mackay2003}), \emph{type II likelihood} (see e.g. \citealp{berger1985}), or \emph{integrated likelihood} (see e.g. \citealp{gneiting2007}). This quantity, which may be interpreted as the prior mean of the likelihood function, will play a central role in Bayesian model uncertainty. Besides computing posterior model probabilities, that have an intuitive interpretation for assessing model uncertainty within the family at hand (see Section \ref{ss:interpretation}), it is also useful to conduct pairwise model comparisons between two models, say $\mathcal{M}_d$ and $\mathcal{M}_{d'}$. This can be done using the $\emph{posterior odds}$ against $\mathcal{M}_{d'}$
\begin{equation}
\frac{p(\mathcal{M}_d|\mathcal{D})}{p(\mathcal{M}_{d'}|\mathcal{D})} = \frac{p(\mathcal{D}|\mathcal{M}_d)} { p(\mathcal{D}|\mathcal{M}_{d'})} \frac{p(\mathcal{M}_d) }{p(\mathcal{M}_{d'})  }.
\end{equation}
Posterior odds involve two terms: the prior odds $p(\mathcal{M}_d)/p(\mathcal{M}_{d'})$ which only depend on the prior distribution over the family of models, and the ratio of marginal likelihoods, 
\begin{equation}
\textup{BF}_{d/d'}=\frac{p(\mathcal{D}|\mathcal{M}_{d})} { p(\mathcal{D}|\mathcal{M}_{d'})},
\end{equation}
called the \emph{Bayes factor}---a term partly coined by Alan Turing during World War II, who called it the ``factor'' \citep{good1979}. The main appeal of the Bayes factor is that, regardless of prior probabilities of models, it provides a good summary of the relative support for $\mathcal{M}_d$ against $\mathcal{M}_d$ provided by the data. Although extremely convenient, this simple interpretation has been subject to much debate (see Section \ref{ss:interpretation}).

For several reasons, it is sometimes convenient to look at logarithmic versions of posterior odds and Bayes factors. For example, the quantity $-2 \log \textup{BF}_{d/d'}$ may be interpreted as a criterion similar to the influential \emph{Akaike information criterion} (AIC, \citealp{akaike1974new}). This rationale gave rise to the popular \emph{Bayesian information criterion} (BIC, \citealp{schwarz1978}) that we discuss in Section \ref{ss:laplace}. Similarly, \citet{zellner1978jeffreys} called $-2 \log (p(\mathcal{M}_{d}|\mathcal{D})/p(\mathcal{M}_{d'}|\mathcal{D}))$ the \emph{Jeffreys-Bayes posterior odds criterion} and compared it to the AIC.

Now that we have a posterior distribution over the family of models, how can we make use of this knowledge of model uncertainty to take decisions?

The first answer is $\emph{Bayesian model selection}$: settling for the model with the largest posterior probability, leading to the choice
\begin{equation}
{d^*} \in \argmax_{d \in \{1,...,{d_{\textup{max}}}\}} p(\mathcal{M}_d|\mathcal{D}).
\end{equation}
This offers a systematized way of choosing a single model within the family, and can be seen as an instance of hypothesis testing. It is worth mentioning that Bayesian model selection was originally described by Jeffreys as an alternative to classical hypothesis tests. For perspectives on the links between the different approaches of testing, see e.g. \citet{berger2003}.

However, when no model truly stands out, it is often better to combine all models (or some of the bests) to grasp more fully the complexity of the data. There comes the second important Bayesian approach of model uncertainty: \emph{Bayesian model averaging} (BMA). BMA allows to borrow strength from all models to conduct better predictions. Specifically, assume that we are interested in a quantity $\Delta$, that has the same meaning in all models. This quantity can be a value that we wish to predict (like the temperature of the Pacific ocean using several forecasting models, as in \citealp{raftery2005}), or a parameter that appears in all models (like the coefficient of a linear regression model). For a more detailed discussion on what it means to have ``the same meaning in all models'', see the discussion of \cite{draper1999} and the rejoinder of the excellent BMA tutorial of \cite{hoeting1999bayesian}. The BMA posterior distribution of $\Delta$ will be its posterior distribution within $\mathcal{M}_\textup{Exp}$,
\begin{equation}
\label{eq:BMApost}
p(\Delta | \mathcal{D}) = \sum_{d=1}^{d_{\textup{max}}} p(\Delta | \mathcal{M}_d,\mathcal{D}) p(\mathcal{M}_d|\mathcal{D}),
\end{equation}
which corresponds to a mixture of all model-specific posteriors. Taking the mean of the BMA posterior gives a natural point estimate for predicting the value of $\Delta$,
\begin{equation}
\label{eq:BMAaverage}
\hat{\Delta} = \mathbb{E}_{\small{\Delta}} (\Delta| \mathcal{D}) =  \sum_{d=1}^{d_{\textup{max}}} \mathbb{E}_\Delta (\Delta | \mathcal{M}_d,\mathcal{D}) p(\mathcal{M}_d|\mathcal{D}).
\end{equation}
Sometimes, the average may not be conducted over all models, but solely over a smaller subfamily, as in Madigan's and Raftery's \citeyearpar{madigan1994} \emph{Occam's window}.

These two popular techniques, which will constitute our main focus, can be embedded within a larger decision theoretic framework. In this context, Bayesian model selection corresponds to the 0-1 loss and BMA corresponds to the squared loss (see e.g. \citealp{bernardo1994}, Section 6.1 or \citealp{clyde2004}, Section 6).


\subsection{Interpretations of Bayesian model uncertainty}
\label{ss:interpretation}
\subsubsection{Interpretation of posterior model probabilities} Contrarily to other techniques that tackle the model uncertainty problem, the Bayesian approach produces easily interpretable results. Indeed, posterior model probabilities are readily understandable, even by non-statisticians, because of their intuitive nature. But what is their precise meaning? Formally, for each $d \in \{1,...,{d_{\textup{max}}}\}$, the quantity $p(\mathcal{M}_d|\mathcal{D})$ is the probability that $\mathcal{M}_d$ is true given the data, given that we accept the prior distributions over models and their parameters, and given that one of the models at hand is actually true. There are several points of this statement that need further description. First, the controversial question of the relevance of the chosen priors raises many concerns, as described in Section \ref{ss:priors}. Second, the assumption that one of the models is actually true is often problematic. Indeed, in most applied cases, it appears overoptimistic to assume that the true data-generating model is contained within the tested family. In particular, in problems coming from social sciences or psychology, it seems clear that the true data-generating mechanism is likely to be beyond the reach of scientists (see e.g. \citealp{gelman2013}). However, reasoning from the perspective that one of the models is true may remain scientifically valid on several grounds. Indeed, most scientific inference is made conditionally on models (models that are usually known to be false) in order to actually conduct science---a striking example is that the notoriously false Newtonian mechanics still flourish today, because it provides a scientifically convenient framework. Rather than conditioning on a single model, Bayesian model uncertainty conditions on a set of models. 

Conditioning on a set of models is perhaps as wrong as conditioning on a single model, but it certainly is more useful. Conditioning a scientific process on a wrong hypothesis is indeed acceptable as long as this hypothesis is powerful or useful.

Or, as famously explained by \cite{box1979}, 
\begin{quote}
	\emph{The law $PV = RT$ relating pressure $P$, volume $V$ and temperature $T$ of an ``ideal'' gas via a constant $R$ is not exactly true for any real gas, but it frequently provides a useful approximation and furthermore its structure is informative since it springs from a physical view of the behavior of gas molecules. For such a model there is no need to ask the question ``Is the model true?''. If ``truth'' is to be the ``whole truth'' the answer must be ``No''. The only question of interest is ``Is the model illuminating and useful?''.}
\end{quote}
Therefore, the (significantly less formal) interpretation of posterior model probabilities that we will adopt is that $p(\mathcal{M}_d|\mathcal{D})$ is \emph{the probability that $\mathcal{M}_d$ is the most useful model within the family at hand}. Actually, a formalization of this interpretation (which gives a precise predictive sense to the ``usefulness'' of a model) in the case where the true model is out of hand---this situation is often referred to as the $\mathcal{M}$-open scenario, as defined by \citet[Section 6.1.2]{bernardo1994}---was provided by Dawid's \citeyearpar{dawid1984} \emph{prequential} (predictive sequential) analysis. For discussions on prequential analysis and related predictive interpretations, see also \citet[Section 3.2]{kass1995}, \citet[Section 7]{gneiting2007}, \citet[Section 5.6]{vehtari2012}, and \citet{fong2020marginal}. Similarly, \cite{germain2016} gave a new and theoretically grounded predictive foundation of Bayesian model uncertainty which gives more support to the interpretation that we advocate here. We present a detailed discussion of these predictive takes on model uncertainty in Section \ref{ss:germain}.

One potential issue when interpreting posterior model probabilities is their potential overconfidence. Indeed, in some cases we may end up with one model having a posterior probability of almost one, even though it does not truly stand out. \cite{oelrich2020bayesian} recently provided an insightful discussion on this issue. In particular, they presented a detailed treatment of the Gaussian linear model from this perspective of overconfidence.

Let us finish this subsection by describing a case of even more severe failure. There will be cases  where \emph{none} of the models $\mathcal{M}_1,...,\mathcal{M}_{d_{\textup{max}}}$ are particularly useful. We should not hope that computing the posterior probabilities  $p(\mathcal{M}_d|\mathcal{D})$ will give us valuable insight in that case. In particular, seeing a posterior probability very close to one will only mean that this model seems less useless than the rest. As warned by Jaynes:
\begin{quote}
\emph{Probability theory can tell
us how our hypothesis fares relative to the alternatives that we have specied; it does not have the
creative imagination to invent new hypotheses for us.}
\end{quote}
This essentially means that Bayesian model uncertainty will work as long as we also acknowledge and study the limits of the models at hand (this is related to the concept of \emph{model criticism} that we briefly evoke in the conclusion of this review).

\subsubsection{Interpretation of Bayes factors} A key asset of Bayes factors is that, contrarily to posterior model probabilities, they do not depend on prior model probabilities (which are often arbitrary). However, this independence comes at the price of a more controversial interpretability. Since the Bayes factor is equal to the ratio of posterior odds to prior odds, it appears natural to consider it as the quantification of the evidence provided by the data in favor of a model---or, as \cite{good1952} called it, the \emph{weight of evidence}. 
This interpretation, which dates back to Jeffreys and Wrinch, was advocated notably by \citet[Section 4.3.3]{berger1985} and \citet{kass1995}. This interpretation was criticized by \citet{lavine1999}, who showed that, rather than seeing a Bayes factor as a measure of support, it was more sensible to interpret it as a measure of the change of support brought about by the data. In their words,
\begin{quote}
	\emph{What the Bayes factor actually measures
		is the change in the odds in favor of the hypothesis when
		going from the prior to the posterior.}
\end{quote}
In a similar fashion, \citet{lindley1997} warned against the use of Bayes factors, and suggested to rather use posterior odds.

\subsection{Specifying prior distributions}

\subsubsection{Model prior probabilities: non-informative priors and the simplicity postulate}

When there is little prior information about the plausibility of different models, it is reasonable to follow Keynes's \citeyearpar[Chapter 4]{keynes1921} principle of indifference and to choose the uniform prior over models $p(\mathcal{M}_d)\propto 1$. In this setting, using posterior model probabilities will be equivalent to using Bayes factors. However, it is often appropriate to seek more sensible priors that translate some form of prior knowledge. For example, in variable selection problems that involve a very large number of variables (e.g. $10.000$ genes), it appears reasonable to give a higher prior probabilities to models that involve only a moderate amount of variables (e.g. preferring a priori a model that involve $100$ genes over one that involves $10.000$). For examples of similar approaches, see \citet{narisetty2014} or \cite{yang2016}. Actually, this rationale was already advocated by \citet[p. 46]{jeffreys1961}:
\begin{quote}
	\emph{All we have to say is that the simpler laws have the greater prior
		probabilities. This is what Wrinch and I called the} simplicity postulate.
\end{quote}
This simplicity postulate is linked to a philosophic principle known as \emph{Occam's razor}, named after the 14th century philosopher and theologian William of Occam. Occam's razor essentially states that, in the absence of strong evidence against it, a simpler hypothesis should be preferred to a more complex one. Actually, Bayesian model uncertainty involves \emph{two} Occam's razors. The first one is precisely the simplicity postulate, and the second one is the fact that, when two models explain the data equally well, the simplest one has a larger marginal likelihood (see Section \ref{ss:bayespen}).

\subsubsection{Parameter prior probabilities and the Jeffreys-Lindley paradox}

\label{ss:priors}

In Bayesian parameter inference, the influence of the prior distribution tends to disappear in the long run (when the number of observations tends to infinity).
A formalization of this argument is the celebrated Bernstein-von Mises theorem (see e.g. \citealp[Chapter 10]{van2000}). This phenomenon is less present when tackling model uncertainty, and poorly chosen prior distributions may lead to disastrous results even in the asymptotic regime. A famous instance of this problem is the \emph{Jeffreys-Lindley paradox}, which essentially states that using improper or very diffuse prior distributions for parameters will lead to selecting the simplest model, regardless of the data. Popularized by \cite{lindley1957}, this paradox had already been pointed out by \cite{jeffreys1939}. It is also known as the \emph{Bartlett paradox} because of Bartlett's \citeyearpar{bartlett1957} early insight on it. For more details, see \cite{spanos2013} or \cite{robert2014} on the epistemological side, and \cite{robert1993} or \cite{villa2017} on the technical side. The main concern about this paradox is that diffuse priors are often chosen as default priors because of their objective nature. Thus, some particular care has to be taken when specifying priors in the presence of model uncertainty. While the use of improper or diffuse priors is generally proscribed for model selection purposes, several approaches have been proposed to bypass this problem. A first simple instance where improper priors may be acceptable is the case where a parameter appears in all models, like the intercept or the noise variance of a linear regression model (see e.g. \citealp[pp. 44, 82]{marin2014}). Another option is to use some form of resampling. First, perform Bayesian inference on a subset of the data using a (potentially improper) prior distribution, then use the obtained posterior as a prior for the rest of the data. This idea is the foundation of the \emph{fractional Bayes factors} of \citet{ohagan1995} and the \emph{intrinsic Bayes factors} of \citet{berger1996}. These techniques share the usual drawbacks of subsampling methods: they are computationally intensive and are inadequate when the number of observations is small. As recently suggested by \citet{dawid2015} and \cite{shao2018}, replacing the likelihood by another scoring rule can also alleviate the dangers of working with improper priors.

Since using improper or diffuse priors is difficult in model uncertainty contexts, it appears necessary to use methods that allow to choose proper priors. Many ways of specifying such priors exist (see e.g. \citealp[Section 18.6]{bayarri2013} for several examples). Among them, an interesting and controversial approach is to leverage the \emph{empirical Bayes} technique. For each model, usually empirical Bayes considers a parametric family of priors $(p(\theta |\mathcal{M}_d,\eta))_{\eta \in E_d}$ and treat $\eta$ as a frequentist parameter to be estimated by the data. Usually, $\eta$ is estimated by \emph{maximum marginal likelihood}
\begin{equation}
\hat{\eta} \in \argmax_{\eta \in E_d} \log p(\mathcal{D}|\mathcal{M}_d,\eta),
\end{equation}
but other estimation procedures (like the method of moments) can be used. The prior $p(\theta |\mathcal{M}_d,\hat{\eta})$ is eventually chosen. While choosing such a data-dependent prior might be disconcerting, it can be seen as an approximation to a fully Bayesian approach that would use a prior distribution for $\eta$ \citep[Section 6.3]{mackay1994}. Moreover, it leads to very good empirical and theoretical performances in several contexts, such as linear regression \citep{cui2008,liang2009,latouche2016}  or principal component analysis \citep{bouveyron2016}. In a sense, the empirical Bayes maximization problem is equivalent to performing continuous model selection by contemplating $E_d$ as the model space. It is sometimes possible to avoid performing maximum marginal likelihood for each model by averaging over all models: this technique is referred to as glocal empirical Bayes \citep{liang2009}.

\begin{figure}
	\centering{
		\includegraphics[width=1\columnwidth]{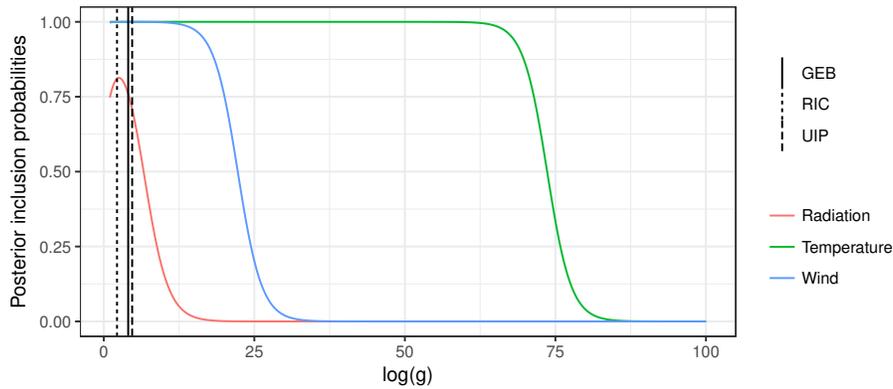}}
	\caption{The Jeffreys-Lindley paradox for linear regression with $g$-priors for the Ozone data set. As $g$ becomes very large, the prior distribution of the regression vector becomes less and less informative, leading to the progressive dismissal of all three explanatory variables. Three proposals of automatic determination of $g$ are also displayed: global empirical Bayes (GEB), risk information criterion (RIC) and unit information prior (UIP, see e.g. \citealp{liang2009}, for more details on these three techniques).}
	\label{fig:jl}
\end{figure}

\paragraph{Example: the Jeffreys-Lindley paradox for predicting ozone concentration} Considering the Ozone data set of \citet{chambers1983}, we wish to predict daily ozone concentration in New York city using three explanatory variables: wind speed, maximum temperature, and solar radiation. For this purpose, we use linear regression with Zellner's \citeyearpar{zellner1986} $g$ prior.  As usual in a variable selection framework, we index the model space using a binary vector $\mathbf{v} \in \{0,1\}^3$ which indicates which variables are deemed relevant in model $\mathcal{M}_\mathbf{v}$. We denote by $\mathbf{Y}$ the vector of observed concentrations, and by $\mathbf{X}$ the matrix of explanatory variables. There are $\# \{0,1\}^3 = 8$ models, defined by
\begin{equation} \label{eq:reglin}
\mathcal{M}_\mathbf{v}: \mathbf{Y}= \mu \mathbf{1}_p + \mathbf{X}_\mathbf{v}\boldsymbol{\beta}_\mathbf{v} + \boldsymbol{\varepsilon}, \end{equation}
where $\boldsymbol{\varepsilon} \sim \mathcal{N}(0,\phi^{-1}\mathbf{I}_n)$. As in \citet{liang2009}, we consider the following prior distributions:
\begin{equation} \label{eq:zellner}
p(\mathcal{M}_\mathbf{v}) \propto 1, \; p(\mu,\phi| \mathcal{M}_\mathbf{v} ) \propto \frac{1}{\phi}, \; \textup{and} \; \boldsymbol{\beta}_\mathbf{v} | \phi,  \mathcal{M}_\mathbf{v}  \sim \mathcal{N}\left(0,\frac{g}{\phi}(\mathbf{X}^T\mathbf{X})^{-1}\right).
\end{equation}
The prior distribution of $\mu$ and $\phi$ is improper, but this is acceptable because both parameters appear in all models (which is not the case for $\boldsymbol{\beta}_\mathbf{v}$). When $g$ becomes large, the prior distribution of $\boldsymbol{\beta}_\mathbf{v}$ becomes very flat, and the Jeffreys-Lindley paradox comes into play. To get a grasp of this phenomenon, we may look at the \emph{posterior inclusion probabilities} for all three variables, defined as the posterior probability that the corresponding coefficient is nonzero (Figure \ref{fig:jl}). When $g$ is very large, Bayesian model uncertainty suggests that all three variables are useless. Three popular ways of automatically choosing $g$ are also displayed. As we can see, using any of these reasonable choices allows to get very far from the Jeffreys-Lindley regime.

While we separated here for clarity the problems of finding priors over model space and parameters, it is worth mentioning that several interesting works considered the \emph{simultaneous} specification of these priors \citep{dawid2011,dellaportas2012}. Let us finish this subsection by quoting \citet{jordan2011}, then president of the International Society for Bayesian Analysis, summarizing a survey he conducted across several senior statisticians regarding important open problems in Bayesian statistics,
\begin{quote}
	\emph{Many people feel that prior specification for model selection is still wide open.}
\end{quote}

\subsection{Theoretical guarantees of Bayesian model uncertainty}

\label{ss:germain}

Theoretical guarantees of model selection schemes can fall within several frameworks. The two main criteria at play are usually \emph{the modeling assumptions} (is there a ``true model'' or not?) and \emph{the nature of the guarantees} (asymptotic or not? predictive of explanatory?).

\subsubsection{Is finding a ``true model'' desirable?} In most practical cases, it appears unrealistic to assume that one model within the available family did actually generate the data. However, this assumption is commonly made when statisticians assess the performance of a model selection scheme. A reason for this is that, in the (overly optimistic) framework where there actually were a true model, we would want a good model selection technique to find it. Or, to quote \cite{liang2009},
\begin{quote}
	\emph{While agreeing
		that no model is ever completely true, many (ourselves included) do feel it is useful to study the behavior of procedures under the assumption of a true model.}
\end{quote}
This ``good behavior in the best case scenario'' framework is sometimes considered pointless as this ``best case scenario'' is too unrealistic (see e.g. \citealp{gelman2013}, or \citealp{spiegelhalter2014}). Although we believe that the true model assumption can be of interest, we will not focus on theoretical results that rely on it in this section.

Note however that, sometimes, the true model assumption can be valid. This is for example the case in physics, for example if one wants to choose between Newtonian gravitation or Einstein's general relativity \citep{jefferys1992}.

\subsubsection{Asymptotics} Traditionally, theoretical model selection guarantees aim at ensuring that, in the long run (when $n$ goes to infinity), the studied technique gives a high probability to the best model. If there is no true model, the  closest model in the Kullback-Leibler sense if often considered. This property is usually called \emph{model selection consistency}. For recent perspectives on the subject, we defer the reader to \cite{liang2009} regarding linear regression, \cite{chatterjee2020} for non independent data, and \cite{walker2004}, \cite{dawid2011}, and \cite{chib2016} for broad reviews. An interestingly growing point of view is the high-dimensional scenario where it is assumed that both the number of variables and the number of observations grow to infinity (see e.g. \citealp{moreno2015}, \citealp{barber2016}).

 In this review, while acknowledging the usefulness of both asymptotics and the true model assumption, we wish to focus on the recent findings of which insure that Bayesian model selection allows to find good models from a predictive perspective, \emph{even in non-asymptotic settings and when the true model is not in the family}.

\subsubsection{Out-of-sample performance: links with cross-validation} 

In this subsection, we assume that the data set is compset of $n$ units: $\mathcal{D} = (x_1,...,x_n)$. In that case, for each model $\mathcal{M}_d$, we can factorise the marginal likelihood as
\begin{align}
p(\mathcal{D}  |\mathcal{M}_d ) &= p(x_1|\mathcal{M}_d) p(x_2|x_1,\mathcal{M}_d)  p(x_3|x_2,x_1,\mathcal{M}_d)...p(x_n|x_{n-1},...,x_1,\mathcal{M}_d) \\
&= \prod_{d=1}^{d_\text{max}} p(x_i|x_{1},...,x_{i-1},\mathcal{M}_d).
\end{align}
This simple factorisation has been used at least since  Dawid's \citeyearpar{dawid1984} \emph{prequential} (predictive sequential) analysis \citep[Section 3.2]{kass1995}. It provides a predictive interpretation of the marginal likelihood. Indeed, for a given observation $x_i$, the factor
\begin{equation}
p(x_i|x_{1},...,x_{i-1},\mathcal{M}_d) = \int_{\Theta_d} p(x_i|x_{1},...,x_{i-1},\theta, \mathcal{M}_d) p(\theta | x_{1},...,x_{i-1}, \mathcal{M}_d)d\theta
\end{equation}
corresponds to the \emph{posterior predictive distribution} of $x_i$ given all previous observations $x_{1},...,x_{i-1}$. This factor can thus be interpreted as a measure of how well $\mathcal{M}_d$ would have predicted $x_i$ if it only had seen $x_{1},...,x_{i-1}$. The marginal likelihood is then the product of all these predictive factors, which gives a first very general predictive interpretation of it.

We said "given all previous observations" in the paragraph above, but there might be no notion of order in the data set at all! In that case, it would seem of bit odd to have a predictive interpretation that depends on the ordering of the data. A refinement of the previous factorisation in this case was provided by \citet[Section 7.2]{gneiting2007} and \cite{fong2020marginal}. We will rephrase it in the rest of this subsection.

To formalise that the data are unordered, we assume that $p(\mathcal{D}|\mathcal{M}_d)$ is \emph{exchangeable}: every reorderings of $\mathcal{D}$ are equiprobable under $\mathcal{M}_d$. Formally, for any permutation $\sigma \in S_n$, we have 
\begin{equation}
p(\mathcal{D}|\mathcal{M}_d) = p(x_1,...,x_n|\mathcal{M}_d) = p(x_{\sigma(1)},...,x_{\sigma(n)}|\mathcal{M}_d).
\end{equation}
This if for example the case when all models assume that $x_1,...,x_n$ are i.i.d. In that case, an interesting bridge with the notion of \emph{cross-validation} can be expressed as follows.
Let us split our data into a training set $x_\text{train}$ and a validation set $x_\text{valid}$. Cross-validation (see e.g. \citealp{arlot2010survey}) is based on the idea of evaluating how well $x_\text{valid}$ can be predicted by a model trained on 
$x_\text{train}$, for several splits of the data. Consider now the following specific random splitting scheme:

 \begin{equation}
 \label{eq:split}
\text{split}:\left\{
\begin{array}{ll}
k \sim \text{Uniform}(\{1,...,n\}) \\
\text{valid} \sim \text{Uniform}(\text{subsets of } \{1,...,n\} \text{ of cardinality } k) \\
\text{train} = \{1,...,n\} \setminus \text{valid}.
\end{array}
\right.
\end{equation}
This corresponds to saying that we choose uniformly at random the number of observations that we leave out of the training data. Once this choice is made, all splits are then equiprobable.
We say that $(\text{train},\text{valid}) \sim \text{split}$ when the data are split according to scheme \eqref{eq:split}.

\begin{theorem}[\citealp{fong2020marginal}, Proposition 2]
The log-marginal likelihood is equal to the average cross-validated log-predictive posterior:
\begin{equation}
\log p(\mathcal{D}|\mathcal{M}_d) = \mathbb{E}_\textup{split} \left[\log p(x_\textup{valid}|x_\textup{train},\mathcal{M}_d))\right].
\end{equation}
\end{theorem}
This means that the marginal likelihood may be seen as a measure of out-of-sample performance. As a caveat, the peculiar nature of the random splitting scheme may be source of concern. Indeed, it seems a bit odd to consider validation sets that can be as large as the whole data set. This point is discussed further by \citet[Section 3.2]{fong2020marginal}. We will discuss in next subsection another remarkable link between the marginal likelihood and out-of-sample performance.

\subsubsection{Out-of-sample performance: PAC guarantees} 

\cite{germain2016} established an important bridge between the (essentially frequentist) PAC-Bayes theory and Bayesian model selection. PAC-Bayes theory, introduced by \cite{shawe1997} and \cite{mcallester1998} and championed by \cite{catoni2007}, aims at finding non-asymptotic \emph{probably approximately correct} (PAC) bounds on the generalization error of a machine learning algorithm. As we will see, the PAC-Bayesian machinery also allows to find bounds on the \emph{predictive log-likelihood} (that is, the likelihood of new, unseen data) of a Bayesian model. 

In the following, we consider a supervised setting where we are dealing with $n$ i.i.d. copies $\mathcal{D}=(X,Y)=\left( x_i,y_i \right)_{i\leq n}\in (\mathcal{X}\times \mathcal{Y})^n$ of a random variable $(x,y) \sim p_{\textup{data}}$. Since it is not assumed that the data-generating model lies within the family, prior model probabilities are assumed to be chosen as scores of prior usefulness of the models, and posterior model probabilities cannot eventually be seen as probabilities that the models are true (see Section \ref{ss:interpretation}).
The predictive log-likelihood function, defined for a given model $\mathcal{M}_d$ as, for all $\theta \in \Theta_d$,
\begin{equation}
\mathcal{L}(\theta|\mathcal{M}_d)= \mathbb{E}_{x,y} [\log p(y|x,\theta, \mathcal{M}_d)],
\end{equation}
will be the quantity of interest, as it allows to asses the out-of-sample performance of a model. We will also look at the BMA predictive log-likelihood, defined as, for all $(\theta_1,...,\theta_{d_{\textup{max}}}) \in \Theta_1\times... \times \Theta_{d_{\textup{max}}}$,
\begin{equation}
\mathcal{L}_\textup{BMA}(\theta_1,...,\theta_{d_{\textup{max}}})= \sum_{d=1}^{d_{\textup{max}}} p(\mathcal{M}_d|X,Y) \mathbb{E}_{x,y} [\log p(y|x,\theta_d, \mathcal{M}_d)].
\end{equation}

Although we will not assume that $p_\textup{data}$ lies within the model family, we need to make assumptions on this distribution in order to bound the predictive likelihood. Following, \cite{germain2016}, we will rely on the following \emph{sub-gamma assumption} stated below. For more details on sub-gamma random variables, see e.g. \citet[Section 2.4]{boucheron2013}.
\begin{definition}[Sub-gamma assumption] A Bayesian model $(p(\cdot|\theta)_{\theta \in \Theta},\pi)$ of some data $\mathcal{D}$ coming from a distribution $p_{\textup{data}}$ satisfies the sub-gamma assumption with variance factor $s^2>0$ and scale parameter $c>0$ if the random variable $\log p(\mathcal{D}|\theta)-\mathbb{E}_{\mathcal{D}}\log p(\mathcal{D}|\theta)$ is a sub-gamma random variable, that is that its moment generating function is upper bounded by the one of a Gamma random variable with shape parameter $s^2/c^2$ and scale parameter $c$.
\end{definition}
The validity of the sub-gamma assumption, deeply linked to the theory of concentration inequalities, depends on the true distribution of the data. However, it is not necessary to assume that this true distribution belongs to the model family. Knowing which models satisfy this assumption is of paramount importance, and should be the subject of future work. \cite{germain2016} showed that the linear regression model, for example, satisfies the sub-gamma assumption.

We can now provide out-of-sample guarantees for Bayesian inference under model uncertainty.
\begin{theorem}[\citealp{germain2016}, Corollary 6]
	If the expanded model satisfies the sub-gamma assumption with variance $s^2>0$ and scale $c>0$, we have, with probability at least $1-2\delta$ over the data-generating distribution,
	\begin{equation} \label{eq:BMA}
	\mathbb{E}_{\theta_1,...,\theta_{d_{\textup{max}}}}[\mathcal{L}_\textup{BMA}(\theta_1,...,\theta_{d_{\textup{max}}}) | \mathcal{D} ]\geq \frac{1}{n} \log\left(  \sum_{d=1}^{d_{\textup{max}}}p(Y|X,\mathcal{M}_d)p(\mathcal{M}_d )\delta \right) - \frac{s^2}{2(1-c)},
	\end{equation}and, for each $d \in \{1,...,{d_{\textup{max}}}\}$,
	\begin{equation} \label{eq:MS}
	\mathbb{E}_{\theta}[\mathcal{L}(\theta|\mathcal{M}_d) | \mathcal{D}] \geq \frac{1}{n} \log\left( p(Y|X,\mathcal{M}_d)p(\mathcal{M}_d )\delta \right) - \frac{s^2}{2(1-c)}.
	\end{equation}  \label{theo:germain}
\end{theorem}
This theorem has two important non-asymptotic implications.
\begin{itemize}
	\item Among the family at hand, \emph{the model with the largest marginal likelihood is the one endowed with the strongest PAC guarantees}. This gives strong theoretical support for the  predictive empirical successes of Bayesian model selection, especially in small-sample scenarios (e.g. \citealp{mackay1992a,murphy2010,celeux2012}) or complex nmodels (e.g. \citealp{van2018learning}).
	\item Since the bound obtained using the BMA posterior is tighter, \emph{BMA has stronger PAC guarantees than the best model of the family}. Again, this explains the well-established empirical result that BMA outperforms model selection from a predictive perspective \citep{hoeting1999bayesian,raftery1996,raftery2005,piironen2016}. Note that several results on the superiority of BMA have been presented in the past, but usually relied on the fact that the quantity of interest was actually distributed according to the BMA posterior \citep{raftery2003}.
\end{itemize}
The BMA bound offers guarantees regarding the model averaged log-likelihood. However, in a forecasting context, it is often seen as more relevant to look at the logarithm of the BMA posterior of the response $p(y|x,\mathcal{D})$, as defined in \eqref{eq:BMApost}. Indeed, this criterion corresponds to the logarithmic score of \citet{good1952}, a strictly proper scoring rule widely used to assess the quality of probabilistic forecasts \citep{gneiting2007}.
Using Jensen's inequality, this quantity can be bounded directly using \eqref{eq:BMA}:
\begin{align*}
\mathbb{E}_{x,y}[\log p(y|x,\mathcal{D})]&=\mathbb{E}_{x,y}\left[\log \left( \sum_{d=1}^{d_{\textup{max}}} p(\mathcal{M}_d|X,Y) \mathbb{E}_{\theta_d} [p(y|x,\theta_d, \mathcal{M}_d) |\mathcal{D}]  \right) \right] \\
&\geq \mathbb{E}_{x,y}\left[\sum_{d=1}^{d_{\textup{max}}} p(\mathcal{M}_d|X,Y)  \log \mathbb{E}_{\theta_d} [p(y|x,\theta_d, \mathcal{M}_d) |\mathcal{D}] \right] \\
&\geq \mathbb{E}_{x,y}\left[\sum_{d=1}^{d_{\textup{max}}}  p(\mathcal{M}_d|X,Y)\mathbb{E}_{\theta_d} [\log p(y|x,\theta_d, \mathcal{M}_d)] \right] \\
&\geq \mathbb{E}_{\theta_1,...,\theta_{d_{\textup{max}}}}[\mathcal{L}_\textup{BMA}(\theta_1,...,\theta_{d_{\textup{max}}}) | \mathcal{D} ].
\end{align*}
This gives a new interpretation to the results of \cite{germain2016}. If we compare all models and BMA using the logarithmic scoring rule, then BMA predictions have stronger guarantees than the model with the largest posterior probability, which has itself stronger guarantees than all other models within the family---a related result was obtained by \cite{madigan1994} under the strong assumption that $y|x$ exactly follows the BMA posterior. Note that these guarantees do \emph{not} imply that BMA is always the best way to aggregate models in a predictive perspective. Other aggregations techniques will sometimes outperform BMA, as illustrated by \cite{yao2018using}.

\paragraph{What about point estimation?} This PAC theorem gives guarantees on the posterior expectation of the predictive log-likelihood. However, it is often of interest to have guarantees about point estimates of $\theta$. For each model $d\in \{1,...,{d_{\textup{max}}}\}$, let us consider the posterior mean $\hat{\theta}_d=\mathbb{E}_{\theta}[\theta | \mathcal{D},\mathcal{M}_d]$, a popular Bayesian point estimate, notably because of its decision theoretic optimality under the squared loss \cite[Section 2.2.4]{berger1985}. If we assume that the log-likelihood function is a concave function of $\theta$, Jensen's inequality implies that \begin{equation} 
\mathbb{E}_{x,y}[\log p(y|x,\hat{\theta}_d, \mathcal{M}_d)] \geq  \mathbb{E}_{\theta,x,y} [\log p(y|x,\theta, \mathcal{M}_d)|\mathcal{D}],
\end{equation}
which means that the predictive likelihood evaluated at $\hat{\theta}_d$ will inherit the good PAC properties of the posterior predictive likelihood bounded in \eqref{eq:MS}. Similarly, BMA forecasts obtained with point estimates satisfy
\begin{align}
\mathbb{E}_{x,y}\left[\log\left( \sum_{d=1}^{d_{\textup{max}}} p(y|x,\hat{\theta}_d,\mathcal{M}_d) p(\mathcal{M}_d|\mathcal{D})\right)\right] &\geq \mathcal{L}_\textup{BMA}(\hat{\theta}_1,...,\hat{\theta}_{d_{\textup{max}}}) \\
&\geq \mathbb{E}_{\theta_1,...,\theta_{d_{\textup{max}}}}[\mathcal{L}_\textup{BMA}(\theta_1,...,\theta_{d_{\textup{max}}}) | \mathcal{D} ].
\end{align}

This theorem offers some strong theoretical insight on why Bayesian model uncertainty works well in a predictive setting, especially in small-sample scenarios. However, one could argue that its merit is merely conceptual. Indeed, the fact that the bounds depend on the data-generating distribution makes them very hard to compute in practice. A good sanity check that was conducted by \citet{germain2016} is to assess the tightness of the bound for some known model. Specifically, they considered the linear regression model (which is sub-gamma for some known scale and variance parameters) and observed that the bound was indeed tight.

Actually, Theorem \ref{theo:germain} also has some important practical applications. Indeed, although the bounds themselves depend on unknown sub-gamma parameters, \emph{the differences between bounds associated with different models can be computed according to their posterior odds}. Indeed, the difference between bounds associated with models $\mathcal{M}_d$ and $\mathcal{M}_{d'}$ is exactly $$\frac{1}{n} \log \left(\frac{p(\mathcal{M}_d|\mathcal{D})}{p(\mathcal{M}_{d'}|\mathcal{D})}\right).$$ In case of a uniform prior probabilities over models, the difference is $n^{-1}\log \textup{BF}_{d/d'}$. This gives a new, predictive, interpretation of the Bayes factors as a measure of evidence in favor of a model. If all bounds are tight, this also gives a good estimate of the generalization gain proposed by a certain model.

An important consequence of this is that it provides a way to quantify the benefits of BMA over model selection. In the discussion on the BMA tutorial of \cite{hoeting1999bayesian}, \cite{draper1999} asked ``what characteristics of a statistical example predict when BMA will lead to large gains?''. While suggesting to perform BMA when the ratio $n/p$ is small, \cite{draper1999} insisted on the need of more refined simple rules that will quantify the relevance of performing BMA over model selection. Such a rule can be derived using the PAC bounds. Indeed, the difference between the PAC bound of the BMA posterior and the one of the model with the largest marginal likelihood is exactly $-(1/n)\max_{d} \log p(Y|X,\mathcal{M}_d)$ which means that the benefits of averaging will be less important when the posterior probability of the best model is close to one. While this consideration is unsurprising, another more important consequence is that $-(1/n)\max_{d} \log p(Y|X,\mathcal{M}_d)$ can be seen as a good estimate of the predictive likelihood gain of performing model averaging. This estimate indicates that, when $n$ is large, model averaging is likely to offer little to no gain over model selection, which can lead to poor results if all models are far away from the data generating mechanism (see e.g. \citealp{yao2018using} for perspectives on this phenomenon).

\begin{figure}
		\vspace{-1.1cm}
	\centering
	\includegraphics[width=\columnwidth]{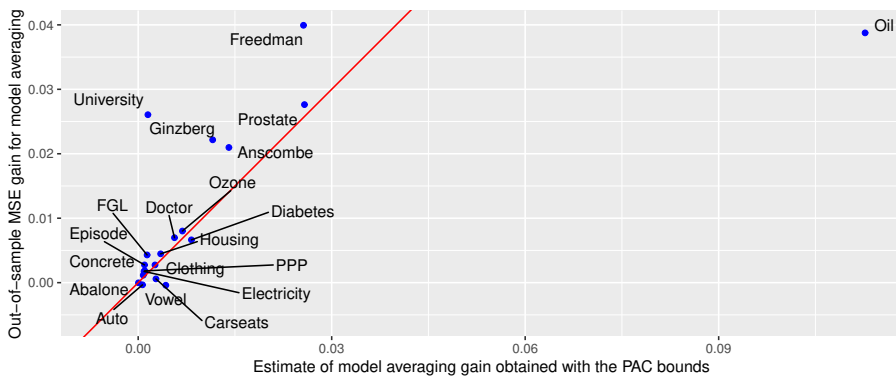}
	\vspace{-2cm}
	\caption{Estimating the BMA gain for 20 linear regression data sets described in Appendix A. Estimate obtained from the PAC bounds versus actual out-of-sample MSE gain obtained with model averaging. Results are averaged over 500 random replications with balanced training/test splits. The identity line, plotted in red, represent the average of what we would get if all data sets satisfied the sub-gamma assumption. Except for the Oil data set, the PAC bounds give a rough but decent estimate of the actual out-of-sample performance. This indicates that, for most of these data sets, the sub-gamma assumption appears quite reasonable. Fitting a robust linear regression model to the 20 data points leads to an estimated slope of 1.54, and an estimated intercept of $-8.86.10^{-4}$. Ignoring the Oil data set leads to a slope of $1.11$ and an intercept of $1.35.10^{-4}$, which is very close to the value suggested by PAC theory (unit slope with no intercept).}
	\vspace{-0.5cm}
	\label{fig:pac}
\end{figure}

\paragraph{Example: how useful is averaging for linear regression?} Consider the Gaussian linear regression model. The usual performance criterion is the mean squared prediction error (MSE), of which the likelihood is the simple affine transformation $\log(2 \pi \sigma) - 1/(2\sigma^2)\textup{MSE}$. According to the PAC bounds, a rough estimate of the \emph{out-of-sample} mean squared error difference between the predictions of the highest probability model and the model averaged ones can be given by $$\textup{MSE(model selection)}-\textup{MSE(BMA)}\approx -(2\hat{\sigma}^2/n)\max_{d \in \{1,...,{d_{\textup{max}}}\}} \log p(Y|X,\mathcal{M}_d)$$ where $\hat{\sigma}$ is an estimate of the residual standard error. We assess the accuracy of this estimate using five standard linear regression data sets (Figure \ref{fig:pac}) and the hyper-$g$-$n$ priors of \cite{liang2009}. Interestingly, this rough estimate  consistently gives a pretty good idea of the gain of performing BMA, and can be seen as a good indicator of whether or not BMA can be useful.

\subsection{Links with penalized model selection}

\label{ss:bayespen}

Both Bayesian and penalty-based approaches build on the likelihood function to perform model selection: while the former \emph{integrates it}, the latter \emph{maximizes it and adds a penalty}. It appears natural to seek foundational connections between these two likelihood treatments. A first step in that direction is to simply remark that the basic identity about conditional probabilities
\begin{equation}
 p(\theta | \mathcal{D} ,\mathcal{M}_d) = \frac{p(\mathcal{D}  |\theta, \mathcal{M}_d) p(\theta |\mathcal{M}_d )}{p(\mathcal{D} |\mathcal{D})},
\end{equation}
implies
\begin{equation}
\label{eq:BMI}
\log p(\mathcal{D} | \mathcal{M}_d) = \log p(\mathcal{D}  |\theta, \mathcal{M}_d) - \log \frac{p(\theta | \mathcal{D} ,\mathcal{M}_d)}{p(\theta | \mathcal{M}_d)},
\end{equation}
for any model $\mathcal{M}_d$ and parameter $\theta \in \Theta_d$ such that $p(\theta | \mathcal{M}_d) \neq 0$. Equation \eqref{eq:BMI}, called by \citet{chib1995marginal} the \emph{basic marginal likelihood identity} means that, for any parameter $\theta$ in the support of the prior distribution, the log-marginal likelihood can be seen as a penalised version of the log-likelihood evaluated at $\theta$. A peculiarity of this formula is that its left hand does not depend on $\theta$, while its right hand does. We are then free to choose whatever $\theta$ we want, so which one should we pick? For several reasons, it makes sense to use a value that has high posterior density, for example a \emph{maximum a posteriori estimate} 
$\theta_{\text{MAP}} \in \argmax_{\theta \in \Theta_d}  p(\theta | \mathcal{D} ,\mathcal{M}_d)$. Indeed, we can expect that in that case the likelihood will be high, and the ratio $p(\theta_{\text{MAP}} | \mathcal{D} ,\mathcal{M}_d) / p(\theta_{\text{MAP}} | \mathcal{M}_d)$ will be larger than one: this means that the penalty term $\log \left( p(\theta_{\text{MAP}} | \mathcal{D} ,\mathcal{M}_d) / p(\theta_{\text{MAP}} | \mathcal{M}_d) \right)$ is likely be positive and indeed act as a penalty. \citet{rougier2020exact} called this penalty the \emph{flexibility}  of model $\mathcal{M}_d$ and argued that it can be interpreted as a measure of model complexity.

It is worth mentioning that the basic marginal likelihood identity \eqref{eq:BMI} has been used both for computing marginal likelihood in practice \citep{chib1995marginal,chib2001marginal} and to study Bayesian model selection asymptotics \citep{chib2016,chatterjee2020}.

Rather than choosing a single $\theta$ in the  basic marginal likelihood identity \eqref{eq:BMI}, one could choose to average other all possible values of $\theta$, this leads to the fertile point of view developed below.

\paragraph{Kullback-Leibler penalization} Another simple penalized view of Bayesian model selection can be derived as follows. Integrating both sides of the basic marginal likelihood identity \eqref{eq:BMI} with respect to the posterior distribution of $\theta$ leads to

\begin{align}
\log p(\mathcal{D}|\mathcal{M}_d)
&=\int_{\Theta_d} p(\theta|\mathcal{D},\mathcal{M}_d) \left(\log p(\theta|\mathcal{D},\mathcal{M}_d) - \log \frac{p(\theta|\mathcal{D},\mathcal{M}_d)}{p(\theta|\mathcal{M}_d)}  \right) d\theta \\
&= \mathbb{E}_{\theta }[\log p(\mathcal{D}|\theta,\mathcal{M}_d) | \mathcal{D}] - \textup{KL}(p(\cdot |\mathcal{M}_d,\mathcal{D})|| p(\cdot|\mathcal{M}_d)).
\end{align}
This means that maximizing the marginal likelihood can be seen as maximizing a penalized version of the posterior mean of the log-likelihood. The penalty term is simply the Kullback-Leibler divergence between the prior and the posterior, and will arguably penalize complex models in a finer way than penalties based on the number of parameters \citep{seeger2003,zhang2006}. Interestingly, this decomposition shows that choosing a too noninformative prior distribution (such as a Gaussian with very large variance) will lead to an explosion of the Kullback-Leibler term, and to overpenalizing the likelihood, thus choosing a perhaps too simple model. This gives an interpretation of the Jeffreys-Lindley paradox described in Section \ref{ss:priors} as an overpenalization phenomenon. Similar model selection schemes based on penalized versions of the posterior mean of the likelihood $\mathbb{E}_{\theta }[\log p(\mathcal{D}|\theta,\mathcal{M}_d)| \mathcal{D} ]$ have been used in the past. Under the general setting
\begin{equation} \label{penalizedsetup}
\textup{score}(\mathcal{D},\mathcal{M}_d)= \mathbb{E}_{\theta }[\log p(\mathcal{D}|\theta,\mathcal{M}_d)| \mathcal{D} ] - \textup{pen}(\mathcal{D},\mathcal{M}_d),
\end{equation}
we have the following correspondances:
\begin{itemize}
	\item $\textup{pen}_{\textup{PBF}}(\mathcal{D},\mathcal{M})=0$ corresponds to the \emph{posterior Bayes factors} of \cite{aitkin1991}.
	\item $\textup{pen}_{\textup{A\&T}}(\mathcal{D},\mathcal{M})=np/2$ corresponds to an estimator of the posterior predictive likelihood proposed by \citet{ando2010}. Note that \citet{ando2010} also proposed a refined criterion that falls within the general setup of \eqref{penalizedsetup}, but whose formula is much more complex.
	\item $\textup{pen}_{\textup{DIC}}(\mathcal{D},\mathcal{M})=\log p(\mathcal{D}|\hat{\theta},\mathcal{M}_d)/2$, where $\hat{\theta}$ is the posterior mean estimate, is equivalent to the \emph{deviance information criterion} (DIC) of \cite{spiegelhalter2002}.
	\item $\textup{pen}_{\textup{WAIC}_1}(\mathcal{D},\mathcal{M}_d)=\log (\mathbb{E}_{\theta} [p(\mathcal{D}|\theta,\mathcal{M}_d) | \mathcal{D} ])/2$ and $$\textup{pen}_{\textup{WAIC}_2}(\mathcal{D},\mathcal{M}_d)=2(\log (\mathbb{E}_{\theta} [p(\mathcal{D}|\theta,\mathcal{M}_d) | \mathcal{D} ])-\mathbb{E}_{\theta }[\log p(\mathcal{D}|\theta,\mathcal{M}_d)| \mathcal{D} ]))$$are equivalent to two versions of the \emph{widely applicable information criterion} (WAIC) of \citet[Section 8.3]{watanabe2009}.
	\item the \emph{Bayesian predictive information criterion} BPIC of \cite{ando2007} uses a complex penalty $\textup{pen}_{\textup{BPIC}}(\mathcal{D},\mathcal{M})$.
\end{itemize}
Several of these frameworks were specifically designed to estimate the posterior mean of the predictive log-likelihood function, which is exactly the quantity bounded by the PAC theorem of \cite{germain2016}. Even though $\textup{pen}_{\textup{BPIC}}$ and $\textup{pen}_{\textup{WAIC}_2}$ lead to asymptotically unbiased estimates of this quantity, the Kullback-Leibler penalty automatically entangled with Bayesian model selection is, to the best of our knowledge, the only framework that provides strong guarantees on small-sample behavior. For more insight on the merits of these various penalization schemes, and their links with cross-validation, see \cite{plummer2008}.

\paragraph{Why is it necessary to penalize the posterior mean of the likelihood?} If we want to maximize the posterior predictive log likelihood, it seems natural to maximize the posterior mean of the log likelihood, which can be seen as an \emph{empirical} estimate of our target. Similarly to the theory of empirical risk minimization (see e.g. \citealp{koltchinskii2011oracle}), it is customary to add a penalty to this empirical estimate to avoid overfitting. From a Bayesian point of view, this necessity can be interpreted as follows. When we compute the posterior mean
\begin{equation}
\mathbb{E}_{\theta }[\log p(\mathcal{D}|\theta,\mathcal{M}_d)| \mathcal{D} ],
\end{equation}
we use the same data \emph{twice} (to find the posterior distribution and to compute the likelihood inside the expectation), which is not consistent with the Bayesian approach. \cite{aitkin1991}, who suggested an unpenalized use of the posterior mean of the likelihood, was criticized by several of his discussants because of this double use of the same data. As explained by \cite{plummer2008}, the penalty is what ``must be paid for using the data (...) twice''.

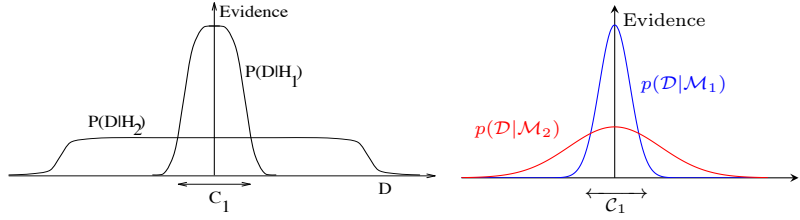
\begin{figure}
	\caption{MacKay's Occam razor plot. \emph{Left:} MacKay's idealized plot, reproduced from \citet[p. 344]{mackay2003}. \emph{Right:} MacKay's plot for a single Gaussian observation.}
	\centering
	\includegraphics[width=0.5\columnwidth]{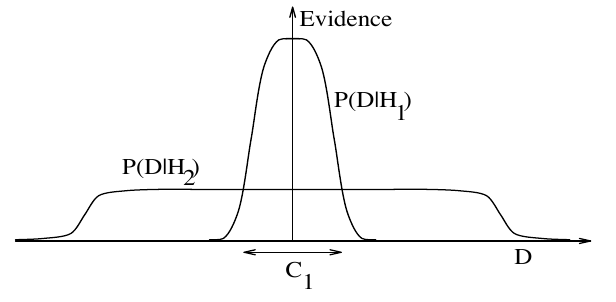}
	\begin{tikzpicture}
	\begin{axis}[every axis plot post/.append style={
		mark=none,domain=-10:10,samples=100,smooth}, 
	axis x line=bottom, 
	axis y line=middle, 
	enlargelimits=upper,
	ylabel=\scriptsize{Evidence}, 
	xlabel=\scriptsize{$\overleftrightarrow{\quad\mathcal{C}_1\quad}$}, 
	x label style={at={(axis description cs:0.46, 0.2)}},
	width=0.5\columnwidth, %
	height=0.32\columnwidth,
	ymax=0.41,
	xtick=\empty,
	ytick=\empty] 
	\addplot {gauss(0,1)}  node [pos=0.56, above right] {\scriptsize$p(\mathcal{D}|\mathcal{M}_1)$};
	\addplot {gauss(0,3)} node [pos=0.35, above left] {\scriptsize$p(\mathcal{D}|\mathcal{M}_2)$};
	\end{axis}
	\end{tikzpicture}
	\label{fig:mackay}
\end{figure}

\paragraph{MacKay's Occam razor interpretation} In his thesis and subsequent work, \cite{mackay1991,mackay1992,mackay2003}, inspired by \cite{gull1988}, drew interesting connections between penalized maximum likelihood methods and Bayesian model uncertainty. The first step is to look at a Laplace approximation of the marginal likelihood. For i.i.d. data, we have, under some (unfortunately not so mild) regularity conditions that we discuss in Section \eqref{ss:watanabe}
\begin{equation} \label{eq:laplace}
\log p(\mathcal{D}|\mathcal{M})= \underbrace{\log p(\mathcal{D}|\hat{\theta},\mathcal{M})}_{\textup{Maximized likelihood}}+ \underbrace{\log p(\hat{\theta}|\mathcal{M}) + \frac{p}{2} \log 2\pi -\frac{1}{2} \log \det A + O_p\left(\frac{1}{n}\right)}_\textup{Occam factor}\end{equation}
where $\hat{\theta}$ is either the maximum a posteriori or the maximum-likelihood estimator of $\theta$ (in the first case the $p \times p$ matrix $A$ is the Hessian of the log posterior, in the latter it is the observed information matrix). This means that, in the long run, \emph{Bayesian model selection is approximatively equivalent to a form of automatically penalized maximum likelihood}. This automatically designed penalty was called the \emph{Occam factor} by \cite{gull1988}. It essentially depends on the prior distribution and on the ``complexity'' of the model. In some simple scenarios like linear regression, the Occam factor can directly be linked to the number of parameters (see e.g. \citealp{latouche2016})---this builds a direct bridge with $\ell_0$ penalization. However, it is not always the case and the Occam factor penalty provides a more sensible regularization than those based on the number of parameters \citep{rasmussen2001}. For a deeper interpretation of the Occam factor penalty, see \citet[p. 349]{mackay2003}. Mackay's other important insight is  a graphical interpretation of this Occam razor effect. Assume for simplicity that there are only two models, one simple ($\mathcal{M}_1$) and one more complex ($\mathcal{M}_2$). The key idea is to plot the marginal distributions of the data $p(\mathcal{D}|\mathcal{M}_d)$ (seen as functions of $\mathcal{D}$) using an idealized unidimensional $\mathcal{D}$-axis where ``simple'' data sets are located near the center of the plot (Figure \ref{fig:mackay}).  On the one hand, the complex model will be able to provide good fits to a larger range of data sets, and the corresponding marginal distribution $p(\mathcal{D}|\mathcal{M}_d)$ will consequently be flatter. On the other hand, the simpler model will concentrate its mass around a limited number of data sets, leading to a more peaky marginal distribution. If the data comes from the $\mathcal{C}_1$ region of MacKay's plot, then the simpler model will have a larger evidence, even though it might not fit the data as well. This illustrates the automatic ``Occam's razor effect'' of Bayesian model uncertainty:  ``more complex'' are more heavily penalised by the marginal likelihood.  Again, here ``more complex'' will not necessarily mean ``with more parameters'', as illustrated by the following recent examples. \citet{duvenaud2016early} showed that the marginal likelihood of Bayesian neural networks could be used to select when to stop training (when training neural networks, it is customary to stop training before convergence to avoid overfitting, a strategy known as \emph{early stopping}). In this example, ``more complex'' models were ``overtrained'' models, and were more penalised by the marginal likelihood. \citet{van2018learning} used the marginal likelihood to select Gaussian process models invariant to transformations of their inputs. In this second example, ``complex'' models mean ``less invariant'' models, and were again penalised by the marginal likelihood.

As described in Section \ref{ss:priors}, another Occam's razor effect can be added by following the simplicity postulate and giving more prior probability to simpler models. 

\paragraph{Example: MacKay's plot for a single Gaussian observation} We propose to plot a simple instance of MacKay's plot (Figure \ref{fig:mackay}). Consider the case where the data consists in a single  Gaussian observation $x \sim \mathcal{N}(\theta^*,1)$ with unit variance. We wish to know whether $\theta^*=0$. The two models are 
$$ \mathcal{M}_1: x \sim \mathcal{N}(0,1),$$ and $$ \mathcal{M}_2: x|\theta \sim \mathcal{N}(\theta,1), \; \theta \sim \mathcal{N}(0,s^2),$$
leading to the marginal distributions
$$x|\mathcal{M}_1 \sim \mathcal{N}(0,1) \; \textup{and} \; x|\mathcal{M}_2 \sim \mathcal{N}(0,1+s^2).$$
The more complex model $\mathcal{M}_2$ will \emph{always} provide a better fit to the data. But if $x$ is small enough, i.e. in the region
\begin{equation}
\mathcal{C}_1=\left[-\left(1-\frac{1}{1+s^2}\right)\log(1+s^2), \left(1-\frac{1}{1+s^2}\right)\log(1+s^2) \right],
\end{equation}
then the simpler zero-mean model will have a larger marginal likelihood. This illustrates the Occam's razor effect. Note that, when $s$ goes to infinity, $\mathcal{C}_1$ becomes infinitely wide, which means that $p(\mathcal{D}|\mathcal{M}_1)$ is everywhere above $p(\mathcal{D}|\mathcal{M}_2)$. In this limiting case, the simpler model will always be preferred: once again, this is an instance of the Jeffreys-Lindley paradox (see Section \ref{ss:priors}).
Another concrete example of MacKay's plot was given (in a discrete setting) by \cite{murray2005}.

\section{Modern practice of Bayesian model uncertainty}

In this section, we review some computational strategies that allow to set Bayesian model uncertainty in motion.

\subsection{Computing marginal likelihoods}

As explained in the previous section, the posterior probabilities of a model $\mathcal{M}_d$ can be computed using its marginal likelihood
\begin{equation} \label{eq:marginal}
p(\mathcal{D}|\mathcal{M}_d) = \int_{\Theta_d} p(\mathcal{D}|\theta,\mathcal{M}_d)p(\theta|\mathcal{M}_d)d\theta.
\end{equation}
This quantity is therefore of paramount importance to account for model uncertainty. Unfortunately, as a potentially high-dimensional integral, it is often very difficult to compute exactly. Several approximation schemes have been developed accordingly. However, closed-form calculation of the marginal likelihood is sometimes feasible. While classical examples include multivariate Gaussian data (see e.g. \citealp{murphy2007}) or linear regression (see e.g. \citealp{bishop2007}, Section 3.5.1, or \citealp{marin2014}, Section 3.4.3), more complex models have also been tackled recently, such as factor analysis \citep{ando2009}, mixtures of independence models \citep{lin2009}, two-sample nonparametric tests \citep{holmes2015}, empirical Bayes principal component analysis \citep{bouveyron2017,bouveyron2016}, and probit regression \citep{durante2019conjugate,fasano2020class}.

\subsection{Markov chain Monte Carlo methods}

Markov chain Monte Carlo methods (MCMC), the Swiss Army knife of modern Bayesian analysis, have been extensively apply to the calculation of marginal likelihoods, posterior odds, or Bayes factors. In this subsection, we simply give pointers to important milestones of an enormous body of work that falls beyond the parsimonious scope of this review paper.

MCMC approaches to model uncertainty may be divided in two families of techniques.

First, \emph{within-model methods} directly attack the marginal likelihoods of models using MCMC. These techniques include notably importance sampling and its variants---in particular Neal's \citeyearpar{neal2001} annealed importance sampling---, nested sampling \citep{skilling2006}, power posteriors \citep{friel2008}, and schemes based on the harmonic mean identity (see e.g. \citealp{weinberg2012}, and references herein). Interesting reviews devoted to this line of work are given by \cite{robert2009b},  \cite{robert2010}, and \cite{friel2012}. A recent important advance in that direction is the scheme of Grosse et al. (2015), who obtained provable upper and lower bounds on the marginal likelihood when the data generating process is known. This allows to evaluate formally the performance of various within-model methods in complex scenarios (see Grosse et al., 2015, for a few examples).
	
The other family of techniques is the one of \emph{transdimensional methods}. Pioneered by \citet{carlin1995} and by Green's \citeyearpar{green1995} reversible jump framework, these approaches aim at obtaining samples from the posterior distribution over both models and parameters. Good reviews are provided by \citet{sisson2005} and \citet{hastie2012}. See also \citet{hee2016} for recent perspectives.

For insightful comparisons between these two approaches, see \citet{chen2000}, \citet{han2001}, and \citet[Section 5]{clyde2004}. A important issue with both approaches is their limited availability in high-dimensional settings. Indeed, in these cases, the parameter space is too vast to be visited properly and MCMC integration becomes more challenging.

\subsection{A little help from asymptotics}

Computing marginal likelihoods, either exactly or using MCMC, is challenging. However, large-sample theory can also provide an interesting guide to build marginal likelihood approximations.

\subsubsection{The Laplace  approximation and BIC}

\label{ss:laplace}

Recall the Laplace approximation of the marginal likelihood of Section \ref{ss:bayespen}:
\begin{equation}
\small  \label{eq:laplace2}
\log p(\mathcal{D}|\mathcal{M}_d)= \log p(\mathcal{D}|\hat{\theta},\mathcal{M}_d)+ \log p(\hat{\theta}|\mathcal{M}_d) + \frac{\dim \Theta_d}{2} \log 2\pi -\frac{1}{2} \log \det A + O_p\left(\frac{1}{n}\right),\end{equation}
where $\hat{\theta}$ is either the maximum a posteriori or the maximum-likelihood estimator of $\theta$ (in the first case the $\dim \Theta_d \times \dim \Theta_d$ matrix $A$ is the Hessian of the log posterior, in the latter it is the observed information matrix, evaluated at $\hat{\theta}$). When this approximation is valid, a $O_p\left(1/n\right)$ approximation of the marginal likelihood can be computed using simply the maximized likelihood and the observed information matrix. Actually, this rationale leads to even simpler approximations. Indeed, approximating the observed information matrix by $n$ times the Fisher information matrix and dropping all the terms that are $O_p(1)$, we end up with
\begin{equation} \label{eq:laplace3}
\log p(\mathcal{D}|\mathcal{M}_d)= \log p(\mathcal{D}|\hat{\theta},\mathcal{M}_d) -\frac{\dim \Theta_d}{2} \log n + O_p\left(1\right).\end{equation}
The crude marginal likelihood proxy $\log p(\mathcal{D}|\hat{\theta},\mathcal{M}_d) -(\dim \Theta_d/2) \log n$ involved in equation \eqref{eq:laplace3} was first derived by \cite{schwarz1978} and extended by \cite{haughton1988}, who also proved that it produces a consistent model selection procedure. From this approximation, an information criterion similar to AIC can be derived, leading to the popular \emph{Bayesian information criterion} (BIC)
\begin{equation}
\textup{BIC}(\mathcal{D},\mathcal{M}_d) = -2 \log p(\mathcal{D}|\hat{\theta},\mathcal{M}_d) +\dim \Theta_d \log n.
\end{equation}
The BIC has the practical advantage that its off-the-shelf expression does not involve the prior distribution whatsoever, at the price of producing a rough $O_p(1)$ approximation of the marginal likelihood. However,  the BIC actually corresponds to an implicit prior distribution. Indeed, assuming that the prior distribution of $\theta$ is a specific data-dependent prior, it is possible to show that Schwarz's proxy actually provides $O_p\left(1/\sqrt{n}\right)$ approximation of the marginal likelihood \citep{kass1995a,raftery1995}. This prior distribution, called the \emph{unit information prior} (UIP), can be interpreted as a weakly informative prior based on an imaginary sample of one observation. For discussions on the merits and dangers of using the UIP or the BIC, see \cite{weakliem1999}, \cite{raftery1999}, and \cite{kuha2004}.

\subsubsection{Towards singular asymptotics}

\label{ss:watanabe}

We remained voluntarily laconic regarding the regularity conditions for the Laplace approximation \eqref{eq:laplace2} to be valid. Several of them are of importance. For thorough theoretical  of these conditions, see \cite{haughton1988} and \cite{kass1990}. We choose here to give details on the conditions that are the most often violated in practice.

First, it is assumed that $\hat{\theta}$ is an interior point of $\Theta_d$. This can be an important issue in many cases (consider for instance a scale parameter, or $g$ in a $g$-prior). Several solutions have been proposed to efficiently tackle this issue \citep{erkanli1994,hsiao1997,pauler1999}.

Moreover, it is assumed that the Fisher information matrix in invertible. This condition is unfortunately violated in non-identifiable models, which are becoming ubiquitous in statistical inference. Such models, often called \emph{singular models}, include mixture models, factor analysis, probabilistic principal component analysis, hidden Markov models, deep neural networks or reduced-rank regression. In these cases, the Laplace approximation is invalid and more refined asymptotic theory has to be invoked. As first exhibited by \citet{watanabe1999}, algebraic geometry proves extremely useful in this context. Specifically, for a wide variety of singular models, a BIC-like approximation was derived by \citet[Theorem 6.7]{watanabe2009},
\begin{equation} \label{eq:watanabe}
\log p(\mathcal{D}|\mathcal{M}_d) = \log p(\mathcal{D}|\hat{\theta},\mathcal{M}_d) - \lambda_d \log n + (m_d - 1) \log \log n  + O_p\left(1\right),
\end{equation}
where $\lambda_d$ is a positive rational number called the \emph{learning coefficient} (also known as the \emph{real log canonical threshold} in the algebraic geometry literature) and $m_d$ is a natural number called the \emph{multiplicity} of $\lambda_d$. For regular models, the learning coefficient is simply equal to $\dim (\Theta_d)/2$ and its multiplicity is one, which means that Watanabe's result reduces to the BIC approximation. However, for singular models, the couple $(\lambda_d,m_d)$ is an often difficult to compute quantity that depends on the true data generating distribution. A major caveat is therefore that, for \eqref{eq:watanabe} to be used, the true model (which is precisely what we are looking for) has to be known beforehand. This would seems to lead to some inextricable circular reasoning problem. To tackle this issue \cite{watanabe2013} proposed to combine his BIC-like approximation with thermodynamic integration (see also \citealp{friel2017}). 
A fully deterministic solution was also provided by \cite{drton2017} who got around the circular reasoning problem by averaging over different learning coefficients. They defined a \emph{singular Bayesian information criterion} (sBIC) as the solution of a well-posed fixed-point problem \cite[Definition 1]{drton2017}. This new criterion has several merits. First, it is a deterministic and computationally cheap $O_p\left(1\right)$ approximation of the marginal likelihood that reduces to the BIC when the model is regular, and is still valid when the model is singular. For these reasons, it can be considered a valid generalization of the BIC.

\paragraph{The singular Occam factor and the predictive power of singular models} Generalizing MacKay's Occam factor rationale described in \eqref{eq:laplace} to singular models leads to the following asymptotic decomposition of the marginal likelihood:
\begin{equation} \label{eq:laplace}
\log p(\mathcal{D}|\mathcal{M}_d)= \underbrace{\log p(\mathcal{D}|\hat{\theta},\mathcal{M}_d)}_{\textup{Maximized likelihood}}+ \underbrace{ (- \lambda_d )\log n + (m_d - 1) \log \log n  + O_p\left(1\right)}_\textup{Occam factor},\end{equation}
which involves the penalty 
$$\textup{pen}_\textup{Occam}(\mathcal{M}_d)= \lambda_d \log n - (m_d - 1) \log \log n.$$
Under relatively mild conditions (see \citealp[Theorem 7.2]{watanabe2009}), it can be shown that the learning coefficient will be a rational number in $[0,\dim (\Theta_d)/2]$, with multiplicity in $\{1,...,\dim \Theta_{d_{\textup{max}}} \}$. Therefore, for singular models, the automatic penalty entangled with Bayesian model selection will be smaller than the regular BIC penalty
$$\textup{pen}_\textup{BIC}(\mathcal{M}_d)=\frac{\dim \Theta_d}{2}\log n.$$
This fact has two interpretations:
\begin{itemize}
	\item For complex models, the number of parameters gives poor insight on the behavior of Bayesian Occam's razor. A phenomenon studied notably by \cite{rasmussen2001}.
	\item Singular models will benefit from their smaller penalties to have potentially larger marginal likelihoods than regular models. Following \cite{germain2016}, let us consider the marginal likelihood as an indicator of predictive performance. In this framework, singular models that fit the data well may therefore have stronger generalization power than regular models in the asymptotic regime. In other words, \emph{singular models may be less prone to overfitting}.
\end{itemize}
The recent empirical successes of deep neural networks constitute perhaps an interesting instance of this phenomenon. In their most common form, deep neural networks \citep{lecun2015,goodfellow2016} are models for supervised learning involving a predictor of the form
\begin{equation}
F(x)=\sigma_1 \circ f_1 \circ \sigma_2 \circ f_2 \circ... \circ f_M(x),
\end{equation}
where $\sigma_1,...,\sigma_M$ are simple nonlinear pointwise functions chosen beforehand, and $f_1,...,f_M$ are learnt affine functions. Empirical evidence strongly suggests that, if the number of observations is very large, using an important number $M$ of layers leads to better generalization performance---state-of-the-art visual recognition systems usually involve hundreds of layers \citep{he2016}. However, this hypothesis still has little theoretical foundation, and the generalization prowesses of deep neural networks remain largely mysterious \citep{zhang2017}. Asymptotic Bayesian model uncertainty provides a heuristic interpretation. While a one-layer network is a regular model, as the number of layers grows, networks become less and less identifiable. Specifically, the Hessian matrix of the log-likelihood of deep networks appears to have many null eigenvalues \citep{sagun2017}, and at a given number of parameters, deeper networks have fewer degrees of freedom in Ye's \citeyearpar{ye1998} sense \citep{gao2016}. It appears therefore reasonable to conjecture that the learning coefficient shrinks when the number of layers grows. If this is true, then, for a given number of parameters, a deeper network will have a higher marginal likelihood provided that there is enough data. This perspective may explain why deep learning resists much more efficiently to overfitting than other more traditional techniques. It is worth noting that using (approximations of) the marginal likelihood as a predictive score has been quite successful in deep learning (see e.g. \citealp{duvenaud2016early}, \citealp{smith2018}, \citealp{khan2019approximate}).

\begin{table}[b!]
	\centering
	\caption{BIC versus sBIC for reduced-rank regression: MSE over 1000 replications.}
	\label{tab:sbic}
	\begin{tabular}{llllll}
		& OLSE         & BIC          & sBIC         & BMA-BIC       & BMA-sBIC               \\ \hline
		eyedata & 10.8 (1.07) & 8.67 (0.536) & 8.67 (0.541) & 8.67 (0.536)  & \textbf{8.60 (0.584)}  \\
		feedstock  & 10.5 (1.72)        &  10.5 (1.53)        &    \textbf{9.79 (1.42) }  &   10.4 (1.52)     & \textbf{   9.79 (1.42)  } \\
		v\'elibs  & 14.9 (0.980) & 16.5 (0.672) & 14.7 (0.624) & 16.4 (0.694) & \textbf{14.5 (0.612)} \\
		&             &              &              &               &                       
	\end{tabular}
\end{table}

\paragraph{Example: BIC versus sBIC for reduced-rank regression} 
Consider the reduced-rank regression framework, as described by \cite{drton2017}. The problem is to linearly predict a multivariate response using some covariate. Each model corresponds to assigning a rank constraint on the regression matrix parameter. Since prediction is the objective, it would appear natural to perform BMA. Given a new covariate value, the BMA estimate of the response is a weighted average of the posterior means obtained for each model. The weights are posterior model probabilities, but are often replaced by BIC-based approximations \citep{hoeting1999bayesian}. However, since this is a singular case, sBIC approximations may be more sensible. To empirically check if this is true, we use three real data sets: ``eyedata'' \citep{scheetz2006}, ``feedstock'' \citep{liebmann2009} and ``v\'elibs'' \citep{bouveyron2015}. To obtain multivariate regression problems, the following preprocessing step was used. The variables were ranked according to the unsupervised feature selection technique of \citet{bouveyron2016}. The first 20 variables were considered as response and the 30 last were considered as covariates. The data are then split equally between training and test set and the performance is assessed (Table \ref{tab:sbic}) using the mean-squared error (MSE). Five estimators are considered: the ordinary least-squares estimator (OLSE) obtained with the full model, OLSEs obtained with models selected by BIC and sBIC, and two BMA estimators. The sBIC-based BMA estimator outperforms all other competitors, illustrating that sBIC provides a more reliable proxy for posterior probabilities than does BIC.

\subsection{Approximate methods for high-dimensional and implicit models}

The last decades have brought about wilder and wilder statistical models. In this subsection, we focus on two kinds of models for which Bayesian model uncertainty is particularly challenging, and has witnessed important advances in recent years: implicit models and high-dimensional models.

\subsubsection{Handling implicit models with likelihood-free inference}

The models that have been studied so far are \emph{explicit} in the sense that, given a parameter value, we have full access to a candidate distribution with density $p(\cdot|\theta,\mathcal{M}_d)$ over the data space, leading to the computation of the likelihood function $\theta \mapsto p(\mathcal{D}|\theta,\mathcal{M}_d)$ which plays a major role within the Bayesian machinery. However, more and more attention is devoted to families of models for which the likelihood function is not available. This context arises when, given a parameter $\theta$, rather than knowing the corresponding candidate distribution $p(\cdot|\theta,\mathcal{M}_d)$, \emph{we are merely able to simulate data from $p(\cdot|\theta,\mathcal{M}_d)$}. Often, the nonavailability of the likelihood comes from the presence of a latent variable that is difficult to integrate. This is for instance the case of popular population genetics models which involve unobserved genealogical histories (see e.g. \citealp{tavare1997}). Other examples include Markov random fields and related models (see e.g. \citealp{stoehr2017}, for a recent review). While the likelihood is extremely hard to compute in these contexts, it also sometimes does not exist whatsoever. This occurs when dealing with \emph{generative adversarial networks} (GANs, \citealp{goodfellow2014}), deep learning models that have vastly improved the state-of-the-art in pseudonatural image generation. 
GANs essentially assume that the data is generated by passing noise through a neural network parametrized by $\theta$. In this case, while it is easy to sample from the distribution of $\mathcal{D} | (\theta,\mathcal{M}_d)$, this distribution has no density \citep{arjovsky2017}, which makes the likelihood not only intractable, but nonexistent.

General-purpose inference within implicit models has been subject to much attention, dating at least back to \cite{diggle1984}. From a Bayesian perspective, the first important contribution came from population genetics with the seminal paper of \citet{tavare1997}, who proposed a scheme for drawing samples from an approximation of the posterior distribution. The fruitful line of work that followed (see e.g. \citealp{csillery2010}, for a review of applications and \citealp{marin2012}, for a methodological overview) has been called \emph{approximate Bayesian computation} (ABC). While parameter inference in implicit models is already extremely challenging, recent efforts have also been concentrated towards accounting for model uncertainty. Model-specific methodologies has led to efficient schemes for estimating the marginal likelihood in several frameworks, such as exponential random graph models \citep{friel2013,bouranis2017}. In a more general setting, several techniques have been proposed to estimate posterior model probabilities using the ABC rationale (see e.g. \citealp{marin2015}, for a review). In particular, \cite{pudlo2015} proposed a scalable approach based on Breiman's \citeyearpar{breiman2001} random forests. Several papers have also tried to apply variational inference to general implicit models \citep{huszar2017,tran2017,tran2017b}. Althought model uncertainty was not the primary focus of these works, such variational approaches lead to the computation of lower-bounds of the marginal likelihood (see next subsection), and can be therefore used to approximate posterior model probabilities.

\subsubsection{Handling high-dimensional models with large-scale deterministic inference}

Families of high-dimensional models combine two major difficulties when accounting for model uncertainty:
\begin{enumerate}
	\item The marginal likelihood of a high-dimensional model $\mathcal{M}_d$, as a $\dim \Theta_d$-dimensional integral, might be extremely difficult to compute, especially using MCMC methods.
	\item Sparse modelling, which is extremely popular in high-dimensional settings because it can lead to increased interpretability and better performance, usually involves a number of candidate models of order $2^p$, where $p$ is the (large) total number of variables. In this setting, it appears impossible to compute posterior probabilities of all models within the family.
\end{enumerate}
We choose to focus here specifically on sparsity, which has arguably constituted the most popular field of statistical research of the last two decades, culminating perhaps with the monograph of \cite{hastie2015} and Cand{\`e}s's \citeyearpar{candes2014} plenary lecture at the International Congress of Mathematicians.

In a sparse modelling context, there is a largest model $\mathcal{M}$ with parameter space $\Theta$ within which all other models are embedded. A convenient way to write models in this context in through the use of binary vectors $\mathbf{v} \in \{0,1\}^{\dim\Theta}$ that can index each model $\mathcal{M}_\mathbf{v}$, such that
\begin{equation}
\Theta_\mathbf{v} = \{ \boldsymbol{\theta} \in \Theta | \textup{Supp}(\boldsymbol{\theta})=\mathbf{v} \}.
\end{equation}
Accounting for model uncertainty now all comes down to studying the posterior distribution of this high-dimensional binary vector $\mathbf{v}$, and model selection can be recast as the following discrete optimization problem
\begin{equation} \label{eq:discrete}
\mathbf{v}^* \in \textup{argmax}_\mathbf{v} p(\mathbf{v}|\mathcal{D}), \; \textup{with} \; \mathbf{v} \in \{0,1\}^{\dim\Theta}.
\end{equation}
Of course, so far, the problem remains exactly as difficult as before, and both the exact posterior of $\mathbf{v}$ and the best model $\mathbf{v}^*$ remain very difficult to compute because of the large number of models. However, using this formalism, we can now make use of the particular \emph{structure} of the model space $\{0,1\}^{\dim\Theta}$ to efficiently approximate  these quantities. There are several ways of building on this structural knowledge to perform approximate but fast model selection. We review here two particularly efficient ones: variational approximations and continuous relaxations.

First, although knowing the exact posterior distribution of $\mathbf{v}$ would require estimating a prohibiting $(2^p - 1)$-dimensional parameter, we can use the binary vector structure to derive a computationally cheaper approximation of the posterior. Specifically, we can consider a \emph{mean-field approximation} $q_\rho(\mathbf{v})$ of the posterior that factorizes as a product of Bernoulli distributions with parameters $\boldsymbol{\rho}=(\rho_1,...,\rho_p)$:
\begin{equation}
p(\mathbf{v} | \mathcal{D}) \approx q_{\boldsymbol{\rho}}(\mathbf{v}) =  \prod_{i=1}^p q_{\rho_i}(v_i)= \prod_{i=1}^p \mathcal{B}(v_i|\rho_i).
\end{equation}
Knowing this approximate posterior distribution conveniently requires to determine only a $p$-dimensional parameter. To insure that the approximation is close to the true posterior, \emph{variational inference} minimizes the Kullback-Leibler divergence between $q_{\boldsymbol{\rho}}(\mathbf{v})$ and $p(\mathbf{v}|\mathcal{D})$. This is equivalent to maximizing a quantity known as the \emph{evidence lower bound} (ELBO)
\begin{equation}
\textup{ELBO}(\boldsymbol{\rho})=\mathbb{E}_{\mathbf{v}\sim q_{\boldsymbol{\rho}}}[\log p(\mathcal{D},\mathbf{v})] + \textup{H}(q_{\boldsymbol{\rho}}),
\end{equation}
with respect to $\boldsymbol{\rho}$. With this approximation, the very challenging computation of all $2^p$ posterior probabilities has been recast as a much simpler continuous $p$-dimensional optimization problem. This idea has been successfully applied to sparse high-dimensional linear and logistic regression \citep{logsdon2010,carbonetto2012,huang2016}. A similar approach, based on a related variational setting called \emph{expectation propagation} \citep{minka2001}, was also used for group-sparse regression \citep{hernandez2013}. For more details on variational inference in general, and notably on optimization strategies for the ELBO, see \citet[Chapter 10]{bishop2007} and \cite{blei2017}.  Beyond the Kullback-Leibler divergence, other divergences have been recently considered for variational inference, such as R\'{e}nyi's $\alpha$-divergences \citep{hernandez2016,li2016} or the $\chi$-divergence \citep{dieng2017}. It is worth mentioning that assessing the quality of the variational approximation $q_{\boldsymbol{\rho}}(\mathbf{v})$ is not an easy task. A way of doing so via importance sampling was recently proposed by \citet{yao2018}.

While variational inference provides a scalable way of tackling the variable selection problem, the mean-field assumption, which states that variable relevances are independent a posteriori, appears quite restrictive, especially when features are very correlated. Another more direct approach to transform the discrete optimization problem into a continuous one is trough making a continuous relaxation and replacing the condition $\mathbf{v} \in \{0,1\}^{\dim\Theta}$ by a continuous constraint $\mathbf{v} \in \mathcal{V} \subset \mathbb{R}^p$. Using the parameter set $\mathcal{V}=\mathbb{R}_+^p$ was the first proposal in that line of work. Introduced in the context of feed-forward neural networks by \cite{mackay1994} and \cite{neal1996} as \emph{automatic relevance determination} (ARD), it led to efficient and sparse high dimensional learning in several contexts, including kernel machines \citep{tipping2001}. Although the original motivation for ARD was mostly heuristic, similarly to the lasso, good theoretical properties were discovered later on \citep{wipf2008,wipf2011}. An approach closer to traditional model selection is the one of \citet{latouche2016}, who used the ARD-like relaxation $ \mathcal{V} = [0,1]^p$ to determine a small subfamily of models over which the marginal likelihood is eventually discretely optimized.  The key advantage of this technique is that while it has the scalability of both the variational approaches and ARD, it still performs exact Bayesian model selection at the end, the only approximation being the fact that only a small subfamily is considered. Beyond supervised problems, such continuous relaxations have also been used to solve high-dimensional sparse PCA problems \citep{archambeau2009,bouveyron2016}.

\paragraph{The ELBO as an approximation of the marginal likelihood} We have seen that the ELBO appears naturally if one wants to approximate a complex posterior using a parametric surrogate that minimizes the Kullback-Leibler divergence. But, as its name suggests, the ELBO also bounds the marginal likelihood (or evidence) and can therefore be seen as an approximation of it. This leads to an approximate procedure to compute posterior model probabilities, which has proven useful in many contexts involving complex posteriors, such as hidden Markov models \citep{watanabe2003}, Gaussian mixture models \citep[Section 10.2.4]{bishop2007} or stochastic block models \citep{latouche2012,latouche2014}. As a non-asymptotic approximation, the ELBO usually compares favourably in small-sample scenarios with the Laplace-like approximations described in Section \ref{ss:laplace}. Moreover, the quality of this approximation may be assessed by also computing \emph{upper bounds} of the marginal likelihood, as suggested by \citet{dieng2017}.

\section{Conclusion}
Bayesian model uncertainty provides a systematized approach of many of the challenges modern statistics has to face: a large number of variables, a potentially low number of observations, and an ever-growing toolset of new statistical models.
As a concluding and tempering note, it is worth reminding and emphasizing that the paradigm of model uncertainty presented in this review has also been subject to much criticism. A philosophical overview of frequentist objections to Bayesian model uncertainty can be found in the thought-provoking monograph of \citet{mayo2018}. Even within the Bayesian community, several lines of work have criticized Jeffreys's framework, both from foundational (e.g. \citealp{gelman2013}) and technical (e.g. \citealp{robert2016}) grounds, leading to alternative paradigms for model uncertainty. Interesting examples of such approaches include the mixture framework of \citet{kamary2014} or methods based on proper scoring rules \citep{dawid2015,shao2018}. We believe that such constructive criticism is vital for Bayesian model uncertainty to help tackle the challenges offered by modern data, keeping in mind that ``the diversity of statistics is one of its strengths'' (Hacking, 1990). In particular, being able to diagnose cases where all models are irrelevant is not possible using model uncertainty, but is precisely the point of \emph{model criticism}, as advocated for example by \citet{gelman2013}. We think that it will be customary in the future to combine model uncertainty with model criticism, in order to design these ``sophisticatedly simple models'' described and desired by \citet{zellner2001}.

\addtocontents{toc}{\protect\setcounter{tocdepth}{-1}}

\section*{Acknowledgement}
This review paper is an extended version of the first chapter of my PhD thesis, which was fuelled by the wonderful energy and guidance of my advisors, Charles Bouveyron and Pierre Latouche. I am also indebted to the members of my PhD committee, whose feedback vastly improved this review: Francis Bach, Gilles Celeux, Julien Chiquet, Nial Friel, Julie Josse, and Jean-Michel Marin.

\section*{Appendix A. Data and software}


\subsection*{Is BMA useful for linear regression?} The data sets used for the BMA experiment of Figure \ref{fig:pac} are taken from various R packages that are listed in the code\footnote{R code for this experiment available at \texttt{https://github.com/pamattei/BMAvsBMS}.}.

\subsection*{BIC versus sBIC for reduced-rank regression}
These results\footnote{R code for this experiment is available at \texttt{https://github.com/pamattei/BMAsBIC}.} presented in Table \ref{tab:sbic} were previously published in my discussion on the paper by \citet{drton2017}.

\bibliography{biblio}

\end{document}